\documentclass[twocolumn,aps,10pt,showpacs,amsmath,superscriptaddress,pra]{revtex4-2}
\usepackage{mathtools}

\usepackage{braket}
\usepackage[utf8]{inputenc}
\usepackage[latin9]{luainputenc}
\usepackage{mathtools}
\usepackage{amsmath}
\usepackage{braket}
\usepackage{breqn}
\usepackage{amssymb}
\usepackage{graphicx}
\usepackage{float}
\usepackage[symbol]{footmisc}
\usepackage[all]{xy}
\usepackage{bm}

\makeatletter

\usepackage{dcolumn}
\usepackage{bm}
\usepackage{ifpdf}
\usepackage{lineno}\usepackage{amsfonts}
\usepackage{bbm}
\usepackage{mathrsfs}
\usepackage{upgreek}
\usepackage{epstopdf}
\usepackage{setspace}
\usepackage{rotating}
\usepackage{multirow}
\usepackage[english]{babel}
\usepackage{color}
\usepackage{hyperref}
\definecolor{med-blue}{RGB}{25,25,112}
\hypersetup{colorlinks, linkcolor={blue},citecolor={blue}, urlcolor={blue}}

\renewcommand{\ket}[1]{\vert{#1}\rangle}

\newcommand\IITM{\,Department of Physics, Indian Institute of Technology Madras, Chennai 600036, India}
\newcommand\CQUICC{\,Center for Quantum Information, Communication and Computing, Indian Institute of Technology Madras, Chennai 600036, India}
\newcommand\TIFRH{\,Tata Institute of Fundamental Research Hyderabad, Gopanpally, Hyderabad 500046, Telangana, India.}
\newcommand\LCN{\,London center for Nanotechnology, UCL, London WC1H0AH, UK}

\makeatother

\begin{document}



\title{Clock-transition-enabled long electron spin coherence of telecom emitters in \texorpdfstring{$\mathbf{CeO}_2$}{CeO2}} 

\author{Basanta Mistri}
\email{basanta@smail.iitm.ac.in}
\affiliation{\IITM}
\affiliation{\CQUICC}
\author{Vishal Ranjan}
\affiliation{\TIFRH}
\author{Siddharth Dhomkar}
\email{sdhomkar@physics.iitm.ac.in}
\affiliation{\IITM}
\affiliation{\CQUICC}
\affiliation{\LCN}
\date{\today}
    

\begin{abstract}
Rare-earth-ion-doped crystals has emerged as a promising platform for quantum technologies, owing to their narrow telecom-band optical emission, long spin memory, and compatibility with silicon integrated photonic architectures. However, the realization of scalable quantum devices requires host materials with intrinsically dilute spin environments to suppress decoherence. In this context, erbium $(\mathrm{Er}^{3+})$ doped cerium oxide $(\mathrm{CeO}_2)$ is an attractive candidate due to the ultra-low concentration of nuclear spins in the host matrix and its compatibility with silicon-based technologies. Here we perform a comprehensive investigation of the coherence properties of $\mathrm{Er}^{3+}$ electron spin qubit in $\mathrm{CeO}_2$ via semiclassical as well as detailed cluster correlation expansion simulations. By systematically exploring magnetic field strength, pulse sequences, erbium concentration, and spin temperature, we identify regimes where decoherence from the spin bath is strongly suppressed. Our investigations illustrate that at dilute doping concentration (of the order of 10 ppb) and sub-Kelvin temperatures, operation near clock transitions enables Hahn-echo coherence times to approach the second timescale even at natural isotopic abundance. Importantly, from a practical standpoint, coherence times on the order of \(\sim 10\) ms are expected even at liquid helium temperature (about 2 K) for similar concentrations. Moreover, we demonstrate that an additional enhancement can be obtained with conventional multi-$\pi$-pulse dynamical decoupling protocols. Thus, our findings establish $\mathrm{Er}^{3+}$ doped $\mathrm{CeO}_2$ as a front-runner for realizing spin qubits, quantum memories, and integrated quantum networks.

\end{abstract}

\maketitle

\section{Introduction}
Rare-earth ion-doped crystals are at the forefront of emerging quantum technologies due to their exceptional optical and spin properties, making them promising platforms for quantum memories~\cite{Probst2015,LeDantec2021,rancic2018coherence,zhong2015optically,Ortu2018}, long-distance quantum communication~\cite{LeDantec2021,rancic2018coherence,zhong2015optically,Ortu2018}, and hybrid quantum systems~\cite{Kinos2021,FernandezGonzalvo2015,Thiel2011,GOLDNER2015,Bertaina2007}. Among these, erbium (\(\mathrm{Er}^{3+}\)) is well suited for telecom-band quantum information applications, as its optical transitions ($\sim$ 1.5~\(\mu\)m) are directly compatible with standard fiber-optic infrastructure~\cite{Saglamyurek2015,rancic2018coherence}. However, harnessing the full potential of \(\mathrm{Er}^{3+}\) electron spins requires maximally extending their coherence times \cite{Berkman2025}, which remains fundamentally limited by interactions with the surrounding spin bath.
Recent advances have demonstrated millisecond scale electron spin coherence (\(T_2\)) ~\cite{rancic2018coherence,LeDantec2021,Welinski2020,Raha2020,Berkman2025} in erbium doped crystals 
through the combined use of techniques such as isotopic purification~\cite{Berkman2025}, dynamical decoupling~\cite{Uysal2025}, and freezing of spectral diffusion at low temperatures via bath spin polarization \cite{Ranjan_PRL_2022,LeDantec2021,RancicPRB}. In addition, engineering of “clock transitions” - magnetic field regimes in which the spin transition frequency is insensitive to first-order fluctuations in the magnetic field~\cite{Marsh2026,rakonjac2020,Ortu2018}, further extending coherence times.
Achieving significantly long coherence in naturally occurring materials without applying complex decoupling sequences, still remains a key bottleneck. Cerium oxide (\(\mathrm{CeO}_2\)), a wide-bandgap oxide with negligible intrinsic nuclear spins \cite{Kanai2022}, is a promising host for erbium quantum emitters. Its crystal structure allows for the incorporation of \(\mathrm{Er}^{3+}\) at ultra-low concentrations \cite{Grant2024,Zhang2024,Seth2025}, granting access to the fundamental limits of decoherence imposed by the magnetic isotope of oxygen (\(^{17}\mathrm{O}\)) present.

Recent experimental studies on $\mathrm{Er}^{3+}:\mathrm{CeO}_2$ have shown that it is a silicon-compatible quantum materials platform, by demonstrating epitaxial thin-film growth on silicon, telecom-band optical measurements, and the estimations of optical and spin coherence properties \cite{Grant2024,Zhang2024,Seth2025}. These works showed that \(\mathrm{Er}^{3+}\) in \(\mathrm{CeO}_2\) can exhibit a narrow optical homogeneous linewidth and long spin-lattice relaxation.
At the same time, the experimentally observed electron-spin coherence times remained comparatively modest, ranging from the sub-\(\mu\)s regime in early measurements to only tens of \(\mu\)s at millikelvin temperatures, and were found to be limited primarily by spectral diffusion, dipolar \(\mathrm{Er}\)--\(\mathrm{Er}\) interactions, and defect-related magnetic noise \cite{Seth2025}. While \(\mathrm{CeO}_2\) provides an exceptionally quiet host from the standpoint of nuclear-spin density, achieving its full coherence potential requires more than material optimization alone. In particular, this motivates the search for operating regimes in which the \(\mathrm{Er}^{3+}\) spin transition is intrinsically protected against magnetic-field fluctuations, such as clock transitions.

In this work, we theoretically investigate the electron spin coherence of \(\mathrm{Er}^{3+}\) in \(\mathrm{CeO}_2\). We show that electron spin Hahn-echo coherence times (\(T_2\)) can approach the second timescale near clock transitions at ultra-dilute doping concentrations (\(\sim 10\) ppb) and sub-kelvin temperatures (10--20 mk). Even at liquid-helium temperatures (\(\sim 2\) K) coherence times on the order of \(\sim 10\) ms are expected at (10--100 ppb) doping concentrations. 
Moreover, we demonstrate that coherence can be further extended through the implementation of dynamical decoupling sequences such as CPMG. 
These results establish erbium-doped \(\mathrm{CeO}_2\) as a leading solid-state platform for long-lived microwave quantum memories and efficient microwave-to-telecom photon transduction within integrated quantum networks.

\section{Material Platform and Spin Hamiltonian}
\label{Sec:II}
\(\mathrm{CeO}_2\) crystallizes in the fluorite structure with cubic \(Fm\bar{3}m\) symmetry, in which each \(\mathrm{Ce}^{4+}\) ion is coordinated by eight oxygen ions and each oxygen occupies a tetrahedral environment formed by four cerium ions. This highly symmetric cubic host distinguishes \(\mathrm{CeO}_2\) from more commonly studied rare-earth hosts such as \(\mathrm{CaWO}_4\) \cite{LeDantec2021,rancic2018coherence,RancicPRB}, which possess lower crystal symmetry and therefore often exhibit stronger orientational anisotropy in their spin properties. In contrast, the cubic environment of \(\mathrm{CeO}_2\) provides a comparatively simple and symmetric crystal field landscape, which is advantageous for identifying and exploiting magnetically protected transitions.

The dominant cerium isotope (\(^{140}\mathrm{Ce}\)) has zero nuclear spin, and the only naturally occurring magnetic isotope in the host is \(^{17}\mathrm{O}\) (\(I = 5/2\), natural abundance \(\sim 0.038\%\)). As a result, the intrinsic nuclear spin bath is exceptionally dilute compared to many oxide and tungstate hosts. This low-spin environment strongly suppresses host-induced magnetic noise, making \(\mathrm{CeO}_2\) a promising platform for long-lived rare-earth spin qubits without requiring isotopic purification.
When erbium is incorporated into $\mathrm{CeO}_2$, it is expected to predominantly occupy cerium lattice sites. Since the host cation is tetravalent ($\mathrm{Ce}^{4+}$) while the dopant is trivalent ($\mathrm{Er}^{3+}$), charge compensation is generally required, most commonly through nearby oxygen vacancies or related local defect complexes~\cite{Grant2024}.

\renewcommand{\figurename}{FIG.}
\begin{figure*}[ht]
    \centering
    \includegraphics[width=1\textwidth]{ 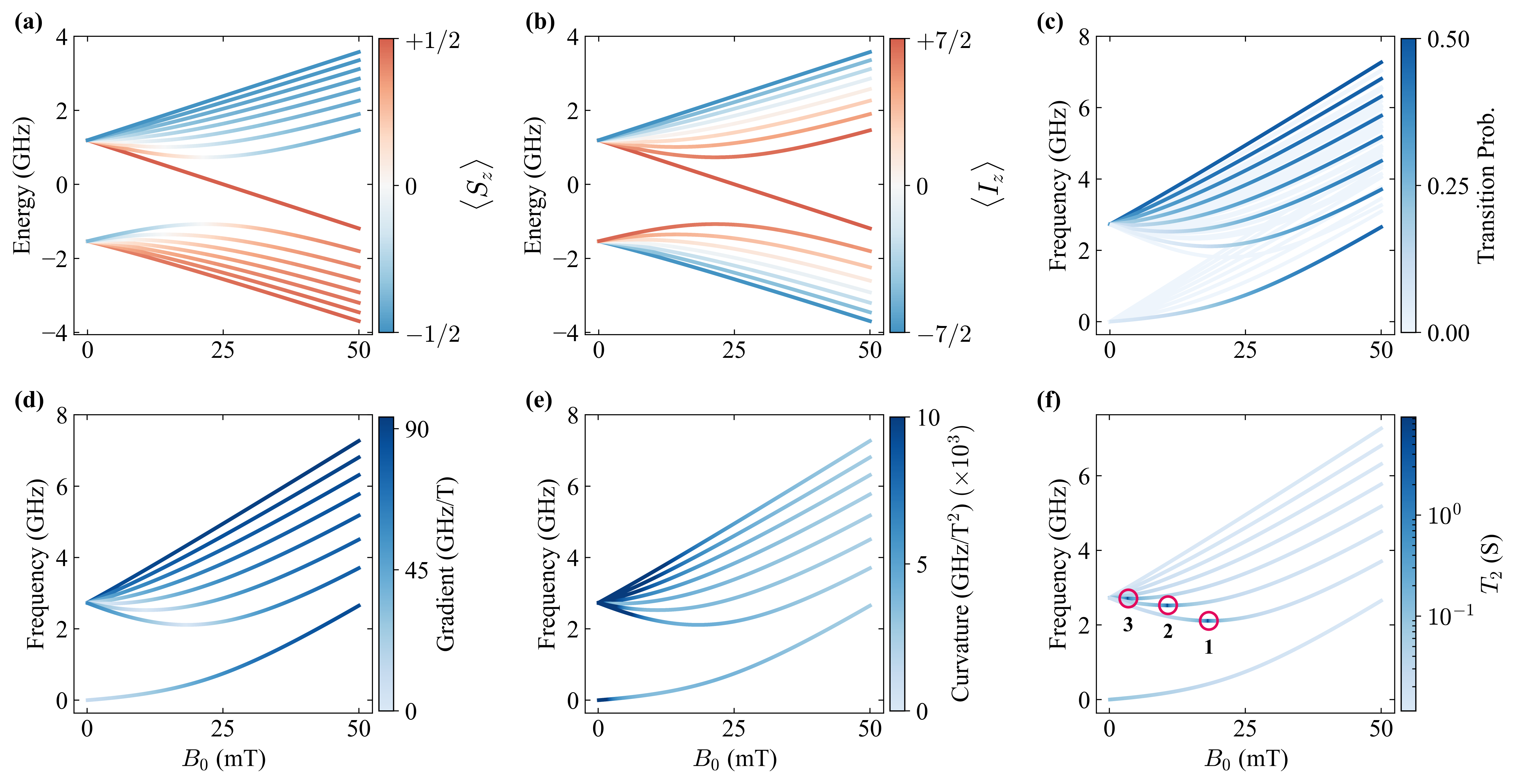}
    \caption{\textbf{Energy-level structure of the central spin system as a function of applied magnetic field $\bm{B_{0}}$.} Energy eigenvalues with color indicating the \textbf{(a)} electronic spin character $\langle S_z \rangle$ of each state, \textbf{(b)} nuclear spin character $\langle I_z \rangle$ of the states. \textbf{(c)} Transition energies between all possible pairs of states, with color-bar representing the transition probabilities. \textbf{(d)},\textbf{(e)} Transition energies as a function of the applied parallel magnetic field $B_0$, with color-bar indicating the magnitude of the transition energy gradient $\partial\nu/\partial B_0$, curvature $\partial^2\nu/\partial B_0^2$ for transitions with significant transition probability. \textbf{(f)} Transition energies as a function of the applied parallel magnetic field $B_0$, color-bar showing calculated Hahn-echo coherence time using rapid $T_2$ estimation methodology (Sec.~\ref{rapid_T2_estimation}). Three distinct transitions exhibiting exceptionally long $T_2$ are identified and highlighted by red circles, labeled as transition ``1, 2, and 3''.}
    \label{fig:1}
\end{figure*}

We focus on the effective spin Hamiltonian of an isolated optically active $\mathrm{Er}^{3+}$ center and its interaction with the surrounding spin bath, which governs the coherence properties studied here. At low temperature, the crystal-field-split ground manifold of \(\mathrm{Er}^{3+}\), can be described as an effective electron spin \(S = 1/2\). For the naturally abundant magnetic isotope \(^{167}\mathrm{Er}\), this electronic degree of freedom is coupled to a nuclear spin \(I = 7/2\), giving rise to a hyperfine-resolved central spin system. In the presence of an external magnetic field \(\mathbf{B}\), the central spin Hamiltonian is written as
\begin{equation}
    \hat{H}_{S} = \mathbf{B} \cdot \bm{\gamma}_e \cdot \mathbf{S}
    + \mathbf{B} \cdot \bm{\gamma}_n \cdot \mathbf{I}
    + \mathbf{S} \cdot \mathbf{A} \cdot \mathbf{I},
\end{equation}
where $\mathbf{S}$ and $\mathbf{I}$ are the electronic spin-$1/2$ and nuclear spin-$7/2$ operators, respectively. Here $\gamma_e \approx 95.3~\mathrm{GHz/T}$ is the effective electronic gyromagnetic ratio, $\gamma_n \approx 1.23~\mathrm{MHz/T}$ is the nuclear gyromagnetic ratio of $^{167}\mathrm{Er}$, and $A \approx 687~\mathrm{MHz}$ is the hyperfine interaction constant. The isotropic form of the hyperfine and $\bm{\gamma}$-tensor reflects the cubic symmetry of the $\mathrm{CeO}_2$ host crystal.

The interplay of Zeeman and hyperfine interactions determines the magnetic-field dependence of the spin eigenstates and transition frequencies. In particular, at low magnetic field, where the Zeeman and hyperfine energy scales are comparable, strong electron-nuclear state mixing gives rise to avoided crossings and magnetically protected transitions, central to the present work.

\section{Simulation Methodology}
\label{Sec:III}
To evaluate the coherence properties of \(\mathrm{Er}^{3+}\) spins in \(\mathrm{CeO}_2\), we use two complementary approaches. First, we perform full spin-bath simulations using CCE, which explicitly captures many-body decoherence arising from the surrounding electron and nuclear spin bath. Second, we use a rapid Hamiltonian-based coherence estimator to efficiently identify magnetic-field regimes where decoherence is expected to be strongly suppressed. Together, these methods allow us to both locate optimal operating points and quantitatively evaluate their coherence performance.
\subsection{Cluster Correlation Expansion (CCE) Bath Simulation}
\label{CCE Method}
To capture full spin-bath-induced decoherence, we perform CCE simulations~\cite{Witzel2006,Yang2008,Yao2006,Rui2012,onizhuk2021pycce}. The total Hamiltonian of the coupled central-spin-plus-bath system is written as
\begin{equation}
    \hat{H} = \hat{H}_{S} + \hat{H}_{SB} + \hat{H}_{B},
\end{equation}
where \(\hat{H}_{S}\) is the central spin Hamiltonian, \(\hat{H}_{SB}\) describes the coupling between the central spin and the bath spins.
$$
 \hat{H}_{SB} = \sum_k \mathbf{S}\cdot \mathbf{A}_{k}\cdot \mathbf{S}_k^B
$$
$S, S_k^B$ are central spin and $k^{th}$ bath spin operator, $A_{k}$ is point-dipole hyperfine coupling tensor between central spin and the $k^{th}$ bath spin.
\(\hat{H}_{B}\) accounts for interactions within the bath itself.
$$
\hat{H}_B = \sum_k \mathbf{B}\cdot\bm{\gamma}_k\cdot\mathbf{S}_k^B \;+\; \sum_{k<l} \mathbf{S}_k^B \cdot \mathbf{D}_{kl} \cdot \mathbf{S}_l^B 
$$
$\gamma_k$ is the gyromagnetic tensor of $k^{th}$ bath spin, $D_{kl}$ is the point-dipole coupling tensor between bath spins $k$ and $l$. 

The bath contains dilute \(\mathrm{Er}^{3+}\) electron spins, treated as effective spin-\(1/2\) fluctuators, and naturally abundant \(^{17}\mathrm{O}\) nuclear spins with \(I = 5/2\), distributed throughout the host lattice. The \(\mathrm{Er}^{3+}\) bath spins are modeled as effective spin-\(1/2\) with an effective electronic \(g\)-factor \((g \approx 6.8)\), consistent with the low-field regime relevant to the present work. 
 Since the magnetic fields considered here remain in the low-field range (up to \(\sim 50\) mT), the fluctuation strength of the dilute nuclear bath remains effectively field independent, and the electron spin bath is only weakly polarized except in the lowest-temperature limit. This makes the present system particularly well suited for combined Hamiltonian-based screening and explicit bath simulation. Within the CCE formalism, the central spin coherence function \(L(t)\) is factorized into contributions from correlated bath-spin clusters,
\begin{equation}
L(t) \approx \prod_C \tilde{L}_C(t)
= \prod_i \tilde{L}_i(t)\prod_{ij}\tilde{L}_{ij}(t)\cdots,
\end{equation}
where \(C\) denotes a particular cluster and \(\tilde{L}_C(t)\) is the corresponding irreducible coherence contribution. The order of the expansion (CCE-1, CCE-2, CCE-3, etc.) is determined by the maximum cluster size included. This framework captures the many-body correlations responsible for nontrivial spin-bath dynamics and enables quantitative simulation of coherence under Ramsey \cite{Ramsey1950}, Hahn-echo \cite{Hahn_1950}, and Carr-Purcell-Meiboom-Gil (CPMG) \cite{Carr_1954, MG1958} sequences. The numerical convergence of the CCE calculations with respect to cluster order, bath radius, and dipolar interaction cutoff has been explicitly verified (see Appendix~\ref{appendix:convergence}), confirming that the reported coherence dynamics are insensitive to truncation of the many-body expansion.

To obtain representative ensemble behavior, we generate statistically independent bath realizations containing both \(^{17}\mathrm{O}\) nuclear spins and \(\mathrm{Er}^{3+}\) electron spins at specified concentrations and temperatures, and average the resulting coherence times over 150 configurations. We adopt a factorized treatment of the combined electron-nuclear spin bath, i.e., we compute the coherence decay due to the dilute \(\mathrm{Er}^{3+}\) electron spin bath, \(L_e(t)\), and the host nuclear spin bath, \(L_n(t)\), separately within the CCE framework. The total coherence is then approximated as
\begin{equation}
L(t) \simeq L_e(t)\, L_n(t),
\end{equation}
which assumes that the dominant decoherence channels arising from electron and nuclear spins are weakly correlated on the timescales of interest. This approximation is well justified in the present system due to the large difference in interaction strengths and dynamical timescales between the sparse electron spin bath and the highly dilute \(^{17}\mathrm{O}\) nuclear spin bath. The effect of temperature is incorporated through the thermal polarization of the \(\mathrm{Er}^{3+}\) electron spin bath. For a given temperature and magnetic field, the equilibrium polarization of each bath spin is sampled according to the Boltzmann distribution, which determines the spin flip-flop processes. Over the temperature range considered here, the nuclear spin bath remains effectively unpolarized owing to its much smaller Zeeman energy.

\subsection{Rapid \texorpdfstring{$\mathbf{T}_2$}{T2} Estimation}
\label{rapid_T2_estimation}
While CCE provides quantitative coherence predictions, it is computationally expensive to apply over a large magnetic-field parameter space. To efficiently identify clock transitions we therefore use a rapid Hamiltonian-based \(T_2\) estimation method based on the magnetic-field dependence of the central spin transition frequencies ~\cite{mistri2025clockfields,Ranjan2020,Wolfowicz2013}

For a given pair of central-spin eigenstates \(|a\rangle\) and \(|b\rangle\), the transition frequency is
\begin{equation}
    \nu(\mathbf{B}) = E_b(\mathbf{B}) - E_a(\mathbf{B}),
\end{equation}

where \(E_a(\mathbf{B})\) and \(E_b(\mathbf{B})\) are the corresponding field-dependent eigenenergies. We evaluate both the first- and second-order magnetic-field derivatives of the transition frequency,
\begin{equation}
    \bm{\nabla}_B \nu = \left. \frac{\partial \nu}{\partial B_i} \right|_{\mathbf{B}_0},
\qquad
\bm{H}_B = \left. \frac{\partial^2 \nu}{\partial B_i \partial B_j} \right|_{\mathbf{B}_0},
\end{equation}

which quantify the linear and nonlinear susceptibility of the transition frequency to field fluctuations.
Clock transitions occur when the first-order magnetic sensitivity is strongly suppressed, i.e. \(\bm{\nabla}_B \nu \rightarrow 0\), such that decoherence is dominated by residual higher-order terms. Assuming a quasi-static Gaussian distribution of local magnetic-field fluctuations with variance \(\sigma_B^2\), the Hahn-echo coherence envelope can be approximated as
\begin{equation}
L(t) = \exp\left[
-\left(
|\bm{\nabla}_B \nu|^2 \sigma_B^2
+ \frac{1}{2}\|\bm{H}_B\|_F \sigma_B^4
+ \cdots
\right)t^2
\right].
\end{equation}

This approach enables rapid mapping of coherence over magnetic-field strength and orientation without requiring full bath simulations at every point. 
An important assumption of this rapid-estimation approach is that the effective fluctuation amplitude \(\sigma_B\) remains approximately constant over the magnetic-field range being scanned. This approximation is valid in the low-field regime relevant to the present work (typically \(\lesssim 50\) mT), particularly when decoherence is dominated by the sparse \({}^{17}\mathrm{O}\) nuclear spin bath (\(I=5/2\)), whose polarization is essentially field independent over this range. 

Given a known coherence time extracted from the CCE simulations at a reference field of $B_{\mathrm{ref}} = 25~\mathrm{mT}$ (yielding $T_2 \approx 39~\mathrm{ms}$ in the nuclear-spin-dominated regime), we extract $\sigma_B$ and use it to estimate the coherence over nearby field strengths and orientations. This provides an efficient means to map optimal operating points in parameter space before carrying out full many-body simulations.

\section{RESULTS AND DISCUSSION}
\renewcommand{\figurename}{FIG.}
\begin{figure*}[ht]
    \centering
    \includegraphics[width=1\textwidth]{ 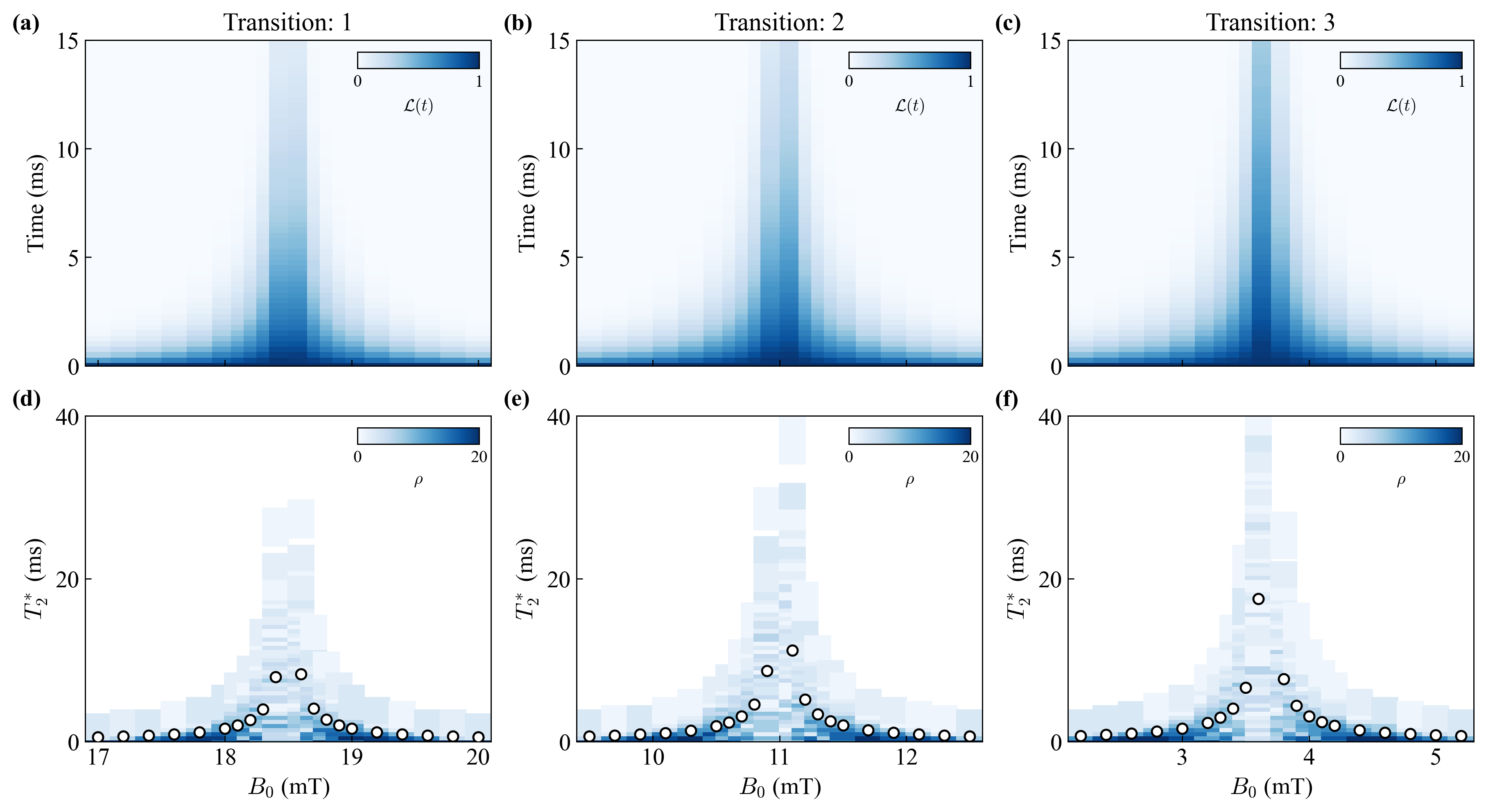}
    \caption{\textbf{Ramsey coherence near clock transitions.} 
\textbf{(a)},\textbf{(b)},\textbf{(c)} The free-induction decay as a function of the magnetic field $B_0$ near the three identified clock transitions. 
\textbf{(d)},\textbf{(e)},\textbf{(f)} The $T_2^*$ heatmap obtained from 150 independent nuclear spin bath configurations, reflecting the statistical variation in dephasing time due to different spin environments. The color corresponds to the number of configurations yielding a given $T_2^*$ at a given magnetic field. The circles indicate the ensemble-averaged coherence time.}
    \label{fig:2}
\end{figure*}

\subsection{ 
 Clock Transitions}
 Figure~\ref{fig:1} presents the calculated energy-level structure of the central spin system as a function of the applied magnetic field $B_0$. Near zero field, the spectrum is dominated by the strong hyperfine interaction between the effective electron spin ($S=1/2$) and the nuclear spin ($I=7/2$). As the magnetic field increases, the Zeeman interaction becomes dominant, driving a crossover from hyperfine-dominated eigenstates to states with well-defined electronic spin projection. In Figure.~\ref{fig:1} (a)-(b) the color-coded energy spectrum represents the expectation value $\langle S_z \rangle$ and $\langle I_z \rangle$, illustrating the electronic and the nuclear character of the eigenstates with increasing $B_0$. The strong electron-nuclear spin mixing gives rise to many possible transitions, not all of which are experimentally relevant. Figure~\ref{fig:1}(c) shows the transition energies between all pairs of eigen-states as a function of $B_0$, with the color scale indicating the corresponding transition probabilities. The dark colored lines isolates the spectrally allowed transitions for the subsequent analysis of magnetic-field insensitivity.
 
Figures~\ref{fig:1}(d) and (e) show the corresponding transition frequency gradients $\partial \nu / \partial B_0$ and curvatures $\partial^2 \nu / \partial B_0^2$. Clock transitions arise at field values where the first-order field dependence is strongly suppressed $\left(\partial \nu / \partial B_0 \rightarrow 0 \right)$, and the residual second-order sensitivity remains minimal, indicating reduced linear and nonlinear susceptibility to magnetic-field noise.
 
Using the rapid $T_2$ estimation framework introduced in Sec.~\ref{rapid_T2_estimation}, we translate the calculated gradients, curvatures, and magnetic field noise amplitude $\left(\sigma_B \approx 0.93 ~\text{nT} \right)$, into predicted Hahn-echo coherence times. Figure~\ref{fig:1}(f) shows the resulting $T_2$ landscape for transitions with significant transition probability. A clear correlation emerges - transitions with reduced magnetic susceptibility, characterized by small gradient and curvature, exhibit substantially enhanced coherence. In particular, near avoided crossings, electron-nuclear hybridization redistributes the spin character of the eigenstates in a way that partially cancels magnetic-field sensitivity, thereby creating optimal coherence points. We identify these at $(B_z, \nu) \approx (18.5~\mathrm{mT},\,2.11~\mathrm{GHz})$, $(10.9~\mathrm{mT},\,2.53~\mathrm{GHz})$, and $(3.6~\mathrm{mT},\,2.71~\mathrm{GHz})$, respectively. For the full CCE calculations, we highlight that the coherence is evaluated at magnetic
fields approximately $2$~G away from the corresponding clock transition positions. This
small offset is chosen because direct CCE convergence becomes increasingly
challenging exactly at the clock condition. Note that this results in slight underestimation of the highest achievable coherence values. Importantly, owing to the cubic symmetry of the host matrix, these optimal points exhibit virtually no dependence on the direction of the magnetic field (see Appendix Figure \ref{fig:angle_sweep} for details), making them highly robust against orientation fluctuations.

All together, this analysis provides a direct route for identifying clock transitions from the spin Hamiltonian alone, without the need for explicit bath simulations. Similar Hamiltonian-based approaches have been successfully employed to extract the decoherence suppressing subspaces in a few material systems \cite{mistri2025clockfields,Ranjan2020,Wolfowicz2013}. That said, aforementioned approach tends to overestimate coherence lifetimes in the vicinity of clock transitions. Therefore, having established these optimal operating regions through rapid estimation, we now turn to a full microscopic treatment of decoherence near the identified clock transitions.
\subsection{Decoherence Dynamics Near Clock Transitions}

\subsubsection{Decoherence Under Ramsey Sequence}
We now investigate the decoherence dynamics near the identified clock transitions under the Ramsey sequence (free-induction decay). At ultralow doping concentration and sub-kelvin temperature, the dominant source of dephasing is the ${}^{17}\mathrm{O}$ nuclear spin bath ($I=5/2$), which generates slowly varying local magnetic fields at the central $\mathrm{Er}^{3+}$ electron spin. 
As Ramsey protocol is directly sensitive to slowly varying Overhauser fields, it provides a direct probe of residual magnetic-field fluctuations.

Figures~\ref{fig:2}(a)-(c) show the calculated Ramsey coherence envelopes $L(t)$ as a function of the magnetic field $B_0$ near the three identified clock transitions, labeled transition 1, transition 2, and transition 3. As the magnetic field approaches the clock field, a pronounced enhancement in coherence is observed. Away from the clock condition, the coherence decays rapidly due to strong linear magnetic-field sensitivity to nuclear spin fluctuations. To quantify the statistical variation arising from different bath realizations, we extract the inhomogeneous dephasing time $T_2^*$ by fitting each Ramsey coherence trace $L(t)$ to a stretched exponential function
\begin{equation}
    L(t)=\exp\left[-\left(\frac{t}{T_2^*}\right)^n\right].
    \label{eq:stretched_decay}
\end{equation}
The resulting distributions are shown in Figures~\ref{fig:2}(d)-(f) as heat maps of $T_2^*$ versus $B_0$. The color scale indicates the number of bath realizations yielding a given coherence time, while the circles denote the ensemble-averaged $T_2^*$ values. For all three transitions, the ensemble-averaged $T_2^*$ exhibits a pronounced maximum near the clock field, confirming that suppression of first-order magnetic susceptibility directly translates into enhanced coherence under realistic many-body bath dynamics. The maximum achievable $T_2^*$ differs among the three transitions reaching $T_2^* \approx 8.3~m\mathrm{s}$, $1.2~m\mathrm{s}$, and $17.5~m\mathrm{s}$ for transitions 1, 2, and 3, respectively, with stretching factor, $n \approx 2.3$, reflecting differences in residual higher-order magnetic sensitivity and state composition near their respective avoided crossings.

 A similar analysis for Hahn-echo coherence lifetimes ($T_2$) is presented in Appendix \ref{appendix:hahn_echo}. Notably, the extracted $T_2$ values show excellent agreement with the results in Figure \ref{fig:1}(f), thereby validating the effectiveness of the rapid search protocol described in \ref{rapid_T2_estimation}.

\subsubsection{Temperature and Concentration Dependence of Coherence}

\begin{figure}[ht]
    \centering
    \includegraphics[width=0.9\columnwidth]{ 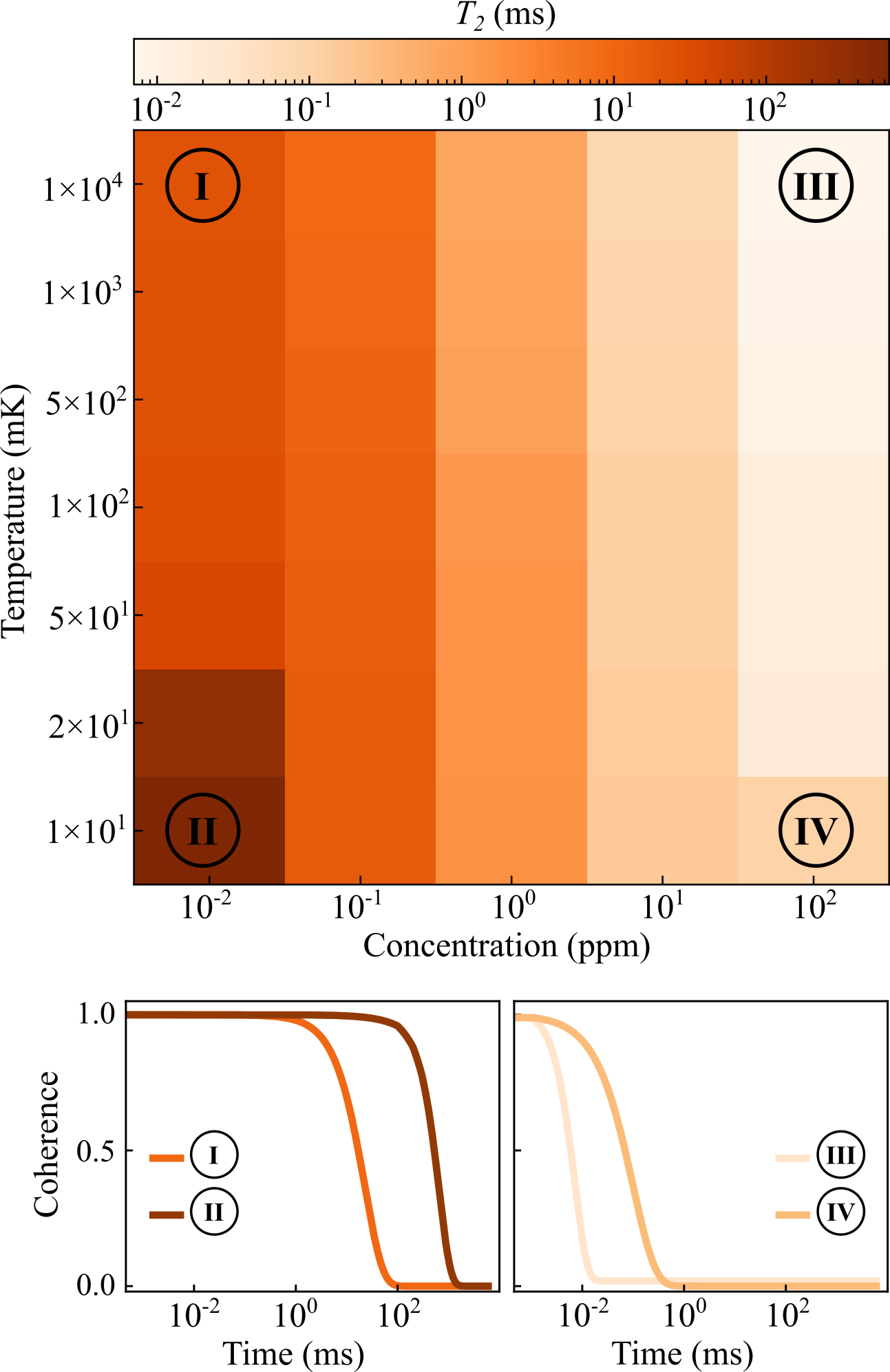}
    \caption{\textbf{Temperature and concentration dependence of Hahn-echo coherence.}
    Top panel: Heatmap of the ensemble-averaged Hahn-echo coherence time $T_2$ as a function of temperature (10 mK–10 K) and $\mathrm{Er}^{3+}$ dopant concentration (10 ppb–100 ppm). 
    Bottom panels show representative Hahn-echo decay curves illustrating the coherence dynamics at four extreme parameter combinations. 
}
    \label{fig:3}
\end{figure}

In addition to the ${}^{17}\mathrm{O}$ nuclear spins, dipolar interactions with neighboring $\mathrm{Er}^{3+}$ electron spins constitute an important decoherence channel, particularly at higher dopant concentrations. Unlike the nuclear spin bath, the electron spin bath is strongly temperature dependent due to thermal polarization at sub-kelvin temperatures. To characterize the influence of these bath degrees of freedom, we analyze the Hahn-echo coherence time $T_2$ as a function of temperature and $\mathrm{Er}^{3+}$ concentration. 
Figure~\ref{fig:3} shows the ensemble-averaged Hahn-echo coherence time $T_2$ as a function of temperature (10 mK - 10 K) and $\mathrm{Er}^{3+}$ concentration (10 ppb - 100 ppm). A clear and systematic trend is observed -- $T_2$ increases as both temperature and concentration are reduced.

At higher concentrations, the density of $\mathrm{Er}^{3+}$ electron spins occupying $\mathrm{Ce}^{4+}$ lattice sites increases, enhancing dipolar flip-flop processes within the electron spin bath. These dynamics generate fluctuating magnetic fields at the central spin, leading to faster decoherence and shorter $T_2$. Moreover, experimentally grown Er$^{3+}$:CeO$_2$ samples may also contain additional paramagnetic defects with effective $g\approx2$, as discussed in Ref.~\cite{Seth2025}. Although, further investigation is necessary to systematically quantify the temperature dependent effect of the paramagnetic spin concentration, a preliminary analysis is provided in Appendix \ref{appendix:effect of g=2 spins}. In particular, we benchmark our results against previously reported data in Ref.~\cite{Seth2025} with additional simulations at 60~mK. The data shown in Appendix Fig.~\ref{fig:g2_bath} is in excellent agreement with both the experimental and numerical results.

\renewcommand{\figurename}{FIG.}

Temperature plays a complementary role. At high temperatures, the $\mathrm{Er}^{3+}$ electron spins are weakly polarized, allowing active spin flip-flop processes that drive spectral diffusion. As the temperature is lowered, the Zeeman splitting exceeds the thermal energy and the bath spins become more polarized. This polarization suppresses flip-flop dynamics and effectively freezes the electron spin bath, thereby reducing magnetic noise. In the sub-kelvin regime, particularly below 100 mK, the bath approaches near-complete polarization, and decoherence from electron spin dynamics is strongly suppressed.

The combined effect of dilution and low temperature produces a dramatic enhancement of coherence. In the ultra-dilute ($\leq 10$ ppb) and millikelvin regime, $T_2$ reaches its maximum values $\left(\approx 641 ~ \text{ms}\right)$, limited primarily by residual nuclear spin interactions and higher-order magnetic sensitivity. In contrast, at concentrations above $\sim 10$ ppm and temperatures above 1 K, electron spin bath fluctuations dominate, leading to rapid coherence decay. Nonetheless, for concentrations in the range of 10-100 ppb, $T_2$ times are expected to be of the order of 10 ms at liquid helium temperatures. This is highly encouraging for single-spin-based quantum transduction and repeater applications, as it eliminates the need for bulky and expensive dilution refrigerators.

\renewcommand{\figurename}{FIG.}
\begin{figure}[ht]
    \centering
    \includegraphics[width=\columnwidth]{ 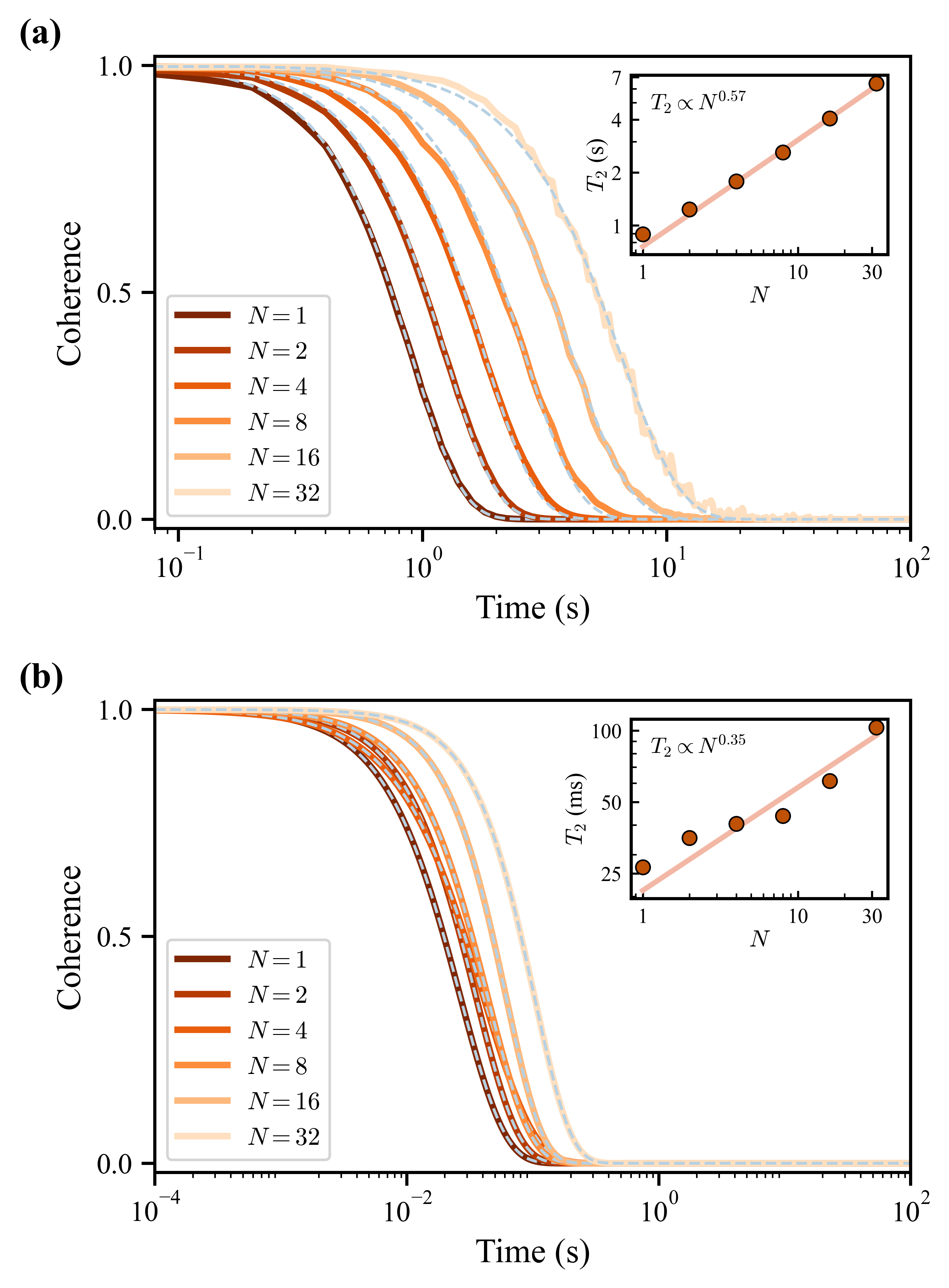}
    \caption{\textbf{Dynamical decoupling enhancement of coherence at the clock transition 3.} \textbf{(a)} Coherence decay $L(t)$ under a CPMG sequence with $N=1$-$32$ $\pi$ pulses for a dilute ensemble (10 ppb erbium) at $T=10$ mK.
    Solid traces show the simulated coherence signals, while dashed curves are fits of the form presented in Equation \ref{eq:stretched_decay}, from which $T_{2}$ and the stretching exponent $n$ are extracted. The inset shows the dependence of $T_{2}$ on pulse number $N$, exhibiting a power-law scaling (see Equation \ref{eq:power_law}), characteristic of nuclear-spin-bath noise filtering. \textbf{(b)} Coherence decay and extracted $T_2$ for the same erbium concentration (10 ppb) but at $T=2$ K.
    }
\label{fig:4}
\end{figure}

\subsubsection{Enhancement of Coherence via CPMG Sequence}
Although operating at low temperature and dilute concentration strongly suppresses spin-bath-induced decoherence, residual fluctuations still limit the achievable coherence time. To further extend coherence, we applied CPMG sequences near the identified clock transition 3. 
Moreover, to assess the influence of the electron spin bath, we compare two representative regimes: (i) $10$ ppb $\mathrm{Er}^{3+}$ at $10$ mK and (ii) $10$ ppb at $2$ K; the results are shown in Figure \ref{fig:4}. At $10$ mK, the $\mathrm{Er}^{3+}$ electron spins are strongly polarized and their flip-flop dynamics are largely suppressed. As a result, decoherence remains predominantly nuclear-spin-limited, and the CPMG sequence yields substantial enhancement of coherence with increasing pulse number $N$. The decay curves are well described by stretched-exponential fits with exponents $n \approx 2$ (see Equation \ref{eq:stretched_decay}) and the extracted coherence times are plotted as a function of number of $\pi$ pulses applied $(N)$ in inset, showing a clear power-law scaling,
\begin{equation}
T_2(N) \propto N^{\eta}.
\label{eq:power_law}
\end{equation}
In this nuclear-spin-dominated regime, the exponent $\eta \approx 0.5$ is consistent with the filtering of slowly fluctuating dipolar nuclear spin noise.
In contrast, at $2$ K, the weakly polarized electron spin bath activates flip-flop processes and generates faster magnetic noise, consistent with the observed stretched-exponential exponents $n \approx 1$--$1.5$. Consequently, the efficiency of dynamical decoupling is reduced, leading to weaker $T_2(N)$ scaling compared to the low-temperature regime, with a reduced exponent $\eta \approx 0.35$. This behavior reflects the broader noise spectrum introduced by the thermally active electron spin bath, which cannot be fully suppressed by a finite number of refocusing pulses. Nevertheless, we emphasize that the expected $T_2$ values approaching 100 ms at liquid helium temperatures represent a highly promising outcome.

\section{Summary \& Outlook}

We have investigated the coherence properties of $\mathrm{Er}^{3+}$ spins in $\mathrm{CeO}_2$ by combining rapid Hamiltonian-based identification of clock transitions with quantitative many-body simulations using CCE. This unified approach enables efficient screening of magnetically protected transitions while retaining predictive accuracy for coherence in a realistic spin environment.

We show that the hyperfine-coupled $\mathrm{Er}^{3+}$ system supports multiple low-field clock transitions arising from electron--nuclear hybridization near avoided crossings. Analysis of first- and second-order magnetic-field sensitivities identifies three transitions with strongly suppressed magnetic noise susceptibility. Rapid $T_2$ estimates indicate substantial intrinsic coherence enhancement at these operating points.

Explicit spin-bath simulations confirm these predictions. Near the identified clock fields, $T_2^*$ is significantly enhanced in the presence of a natural-abundance ${}^{17}\mathrm{O}$ nuclear bath, demonstrating the robustness of magnetic protection. The $\mathrm{Er}^{3+}$ electron spin bath introduces an additional, strongly temperature- and concentration-dependent decoherence channel. In the ultra-dilute (10--100~ppb) and low-temperature regime, thermal polarization suppresses electron spin dynamics, and coherence becomes limited primarily by the nuclear spin bath and residual higher-order magnetic sensitivity.

Dynamical decoupling further extends coherence beyond this limit. Near clock transitions, CPMG sequences suppress low-frequency bath fluctuations and yield a clear power-law scaling of $T_2$ with pulse number. This enhancement is most effective in the nuclear-spin-dominated regime and diminishes as thermally activated electron spin fluctuations broaden the noise spectrum.

These results establish $\mathrm{Er}^{3+}$:$\mathrm{CeO}_2$ as a promising platform for long-lived, telecom-compatible spin qubits. Importantly, millisecond-scale Hahn-echo coherence is expected at liquid-helium temperatures for dopant concentrations in the 10--100~ppb range, eliminating the need for dilution refrigeration. Combined with the relatively weak nuclear spin background, this enables long coherence without isotopic purification, substantially reducing materials complexity.


\section{Acknowledgments}
S.D. thanks the Indian Institute of Technology, Madras, India, and the Science and Engineering Research Board (SERB Grant No. SRG/2023/000322), India, for start-up funding. S.D. and B.M. acknowledge the use of computational facilities supported by a grant from the Mphasis F1 Foundation given to the Center for Quantum Information, Communication, and Computing (CQuICC). V.R. thanks the support from DAE India (RTI 40007). We further acknowledge National Supercomputing Mission (NSM) for providing computing resources of “PARAM RUDRA” at the P.G. Senapathy Center For Computer Resources, Play Field Avenue, Indian Institute of Technology Madras, Tamil Nadu 600036, which is implemented by C-DAC and supported by the Ministry of Electronics and Information Technology (MeitY) and Department of Science and Technology (DST), Government of India.

\section{Code and Data Availability}
The code for Hamiltonian diagonalization as well as gradient and curvature evaluation is available here \cite{github_code_hamiltonian}. Additional code and data are available from the authors upon reasonable request.

\appendix
\section{Convergence Tests for the CCE Calculations}
\label{appendix:convergence}

\begin{figure*}[htp]
    \centering
    \includegraphics[width=1\textwidth]{ 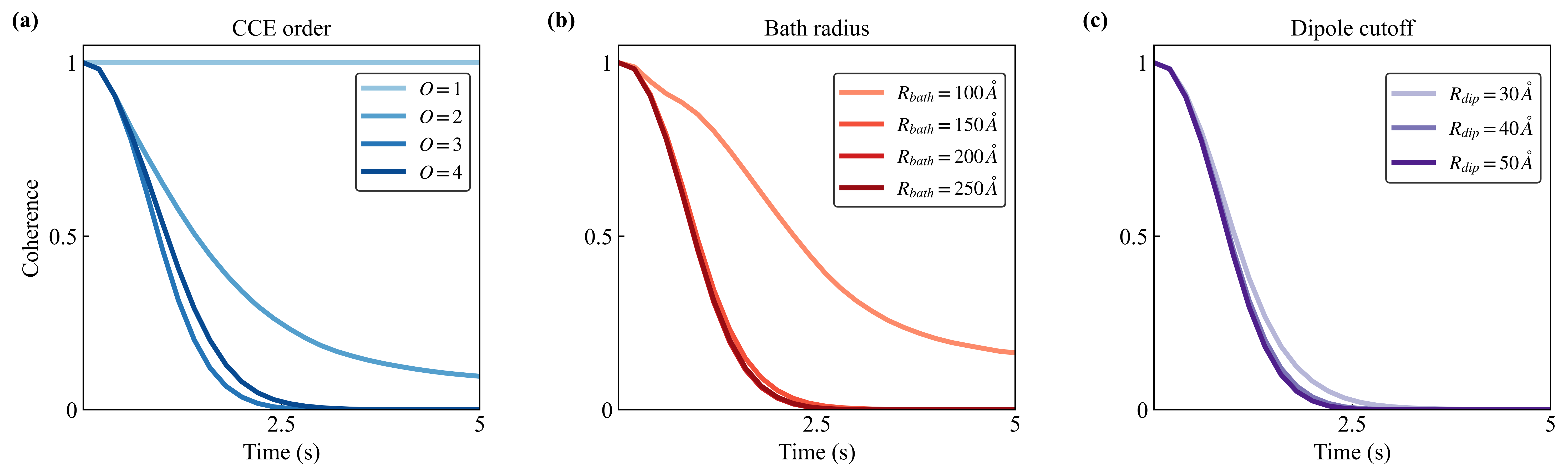}
    \caption{\textbf{Convergence tests for the CCE calculations.}
Hahn-echo coherence decay curves to test numerical convergence with respect to key simulation parameters. 
\textbf{(a)} Dependence on the CCE order $(O)$. Convergence of the coherence dynamics is achieved by $O=3$, indicating that higher-order spin clusters contribute negligibly. 
\textbf{(b)} Dependence on the bath radius $R_{\mathrm{bath}}$, which determines the spatial extent of bath spins included in the simulation. Increasing $R_{\mathrm{bath}}$ beyond $200\,\text{\AA}$ produces negligible changes in the decay, demonstrating convergence with respect to bath size. 
\textbf{(c)} Dependence on the dipolar interaction cutoff distance $R_{\mathrm{dip}}$, which limits pairwise dipolar couplings included in the cluster construction. The coherence curves show minimal variation for cutoffs above $\sim 40\,\text{\AA}$, confirming that the simulation results are insensitive to longer-range interactions. 
These tests establish the parameter choices used in the subsequent CCE calculations of Hahn-echo coherence for nuclear spin dominated bath.}
\label{fig:nspin_conv}
\end{figure*}

To ensure that the coherence dynamics reported in this work are not influenced by numerical truncation, we performed systematic convergence tests for CCE simulations with respect to the principal computational parameters: the CCE order $O$, the bath radius $R_{\mathrm{bath}}$, and the dipolar interaction cutoff $R_{\mathrm{dip}}$. The results are summarized in Figure.~\ref{fig:nspin_conv} for a representative Hahn-echo calculation in the nuclear-spin-dominated regime.

Figure~\ref{fig:nspin_conv}(a) shows the Hahn-echo coherence decay as a function of the CCE order. At first order ($O=1$), the coherence remains nearly unchanged, indicating that independent single-spin contributions alone are insufficient to capture the observed decoherence. Including pair correlations ($O=2$) produces substantial decay, demonstrating that correlated bath dynamics are essential. The results for $O=3$ and $O=4$ are nearly indistinguishable, indicating that the coherence dynamics are well converged by third order and that higher-order bath-spin clusters contribute negligibly under the present conditions. Accordingly, all subsequent simulations were performed using CCE-3.
Figure~\ref{fig:nspin_conv}(b) examines convergence with respect to the bath radius $R_{\mathrm{bath}}$, which sets the spatial extent of bath spins included in the simulation. For small bath sizes, the coherence decay is artificially slow because distant spins that contribute weak but collectively important dipolar fields are omitted. As $R_{\mathrm{bath}}$ is increased, the coherence converges rapidly, with only negligible differences between $200$ and $250$~\AA. This establishes that a bath radius of $R_{\mathrm{bath}} = 200$~\AA\ is sufficient to capture the relevant spin environment.
Figure~\ref{fig:nspin_conv}(c) shows the dependence on the dipolar cutoff distance $R_{\mathrm{dip}}$, which determines the maximum pairwise separation used when constructing interacting spin clusters. The coherence curves for $R_{\mathrm{dip}} = 40$ and $50$~\AA\ are nearly identical, while only a modest deviation is observed for $30$~\AA. This indicates that longer-range dipolar couplings beyond $\sim 40$~\AA\ do not materially affect the calculated coherence dynamics. We therefore adopt $R_{\mathrm{dip}} = 40$~\AA\ in the main simulations as an efficient and well-converged choice.

Taken together, these tests confirm that the CCE parameters used throughout this work are numerically converged and that the reported coherence behavior is governed by the underlying spin physics rather than by finite-cluster or finite-bath truncation effects.

\renewcommand{\figurename}{FIG.}
\begin{figure*}[htp]
    \centering
    \includegraphics[width=1\textwidth]{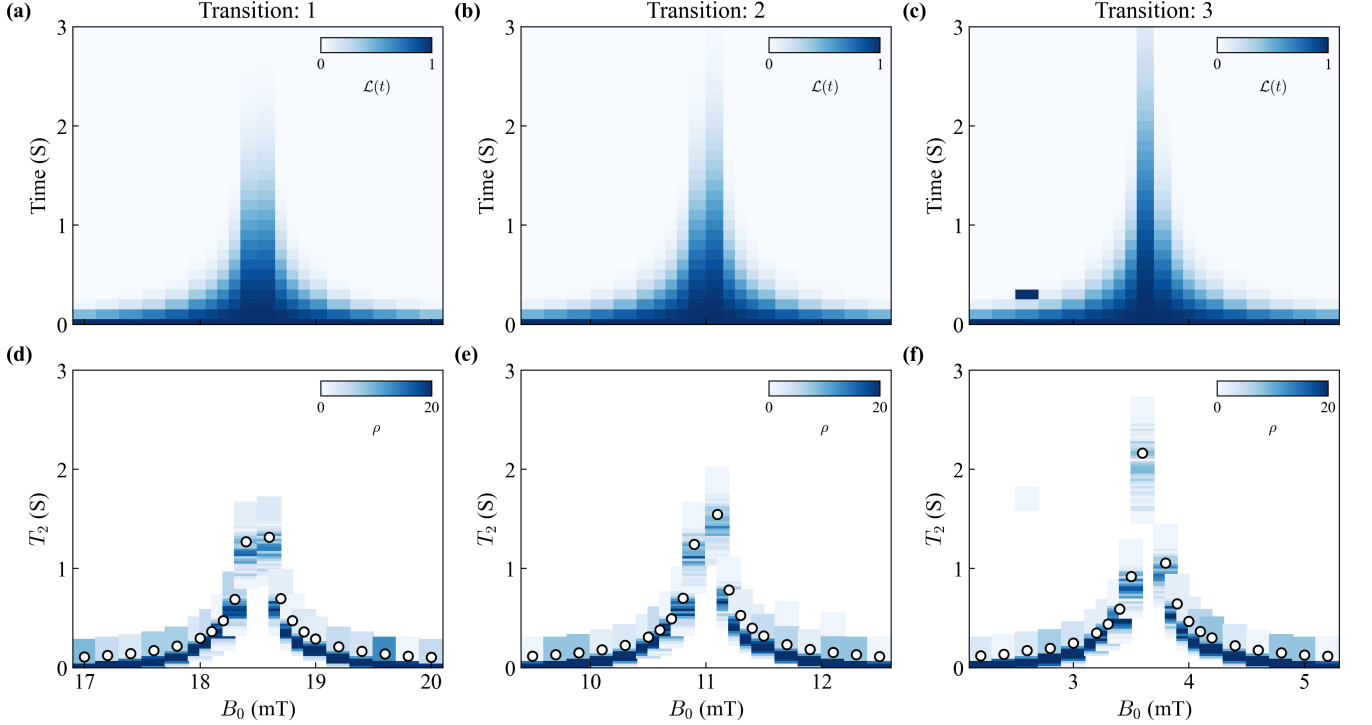}
    \caption{\textbf{Hahn echo coherence near clock transitions}. \textbf{(a)},\textbf{(b)},\textbf{(c)} The Hahn-echo coherence decay as a function of magnetic field $B_0$ near the three identified clock transitions. \textbf{(d)},\textbf{(e)},\textbf{(f)} The $T_2$ heatmap is obtained for 150 independent nuclear spin bath configurations, reflecting statistical variation in $T_2$ due to different spin environments. The color corresponds to the number of configurations with a given $T_2$ at the given magnetic field. The circles indicate the ensemble averaged coherence time.
    }
    \label{fig:Hahn_Echo}
\end{figure*}

\renewcommand{\figurename}{FIG.}
\begin{figure*}[ht]
    \centering
    \includegraphics[width=0.95\textwidth]{ 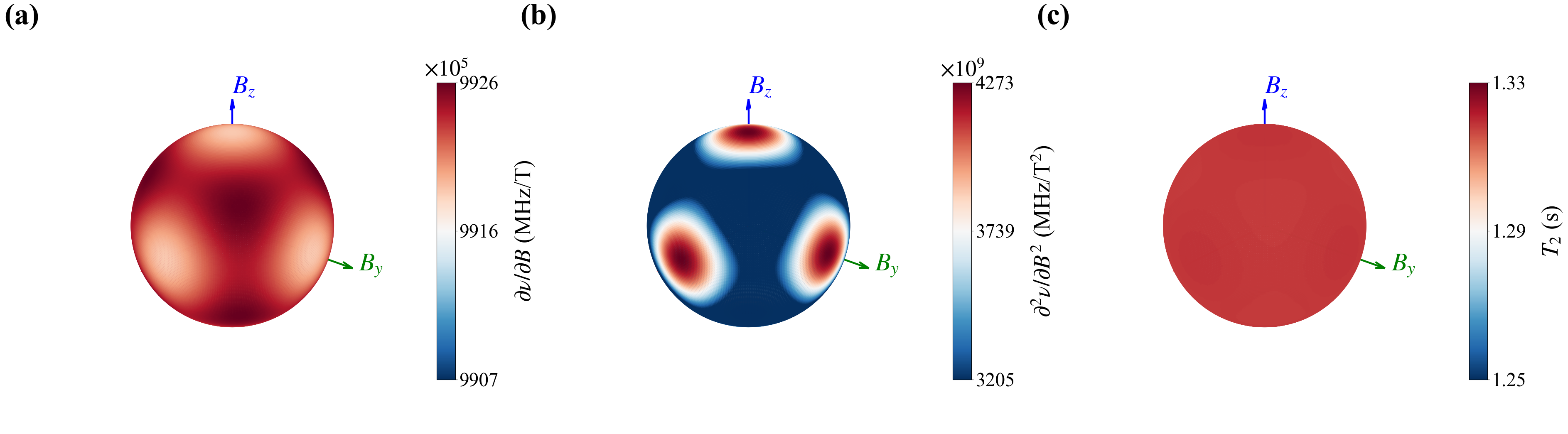}
\caption{\textbf{Angular dependence of magnetic-field sensitivity near a clock transition.}
Spherical maps showing the dependence of coherence properties on the orientation of the applied magnetic field near a clock transition. 
\textbf{(a)} First-order magnetic-field sensitivity (energy gradient) $\partial \nu/\partial B$. 
\textbf{(b)} Second-order magnetic-field sensitivity (curvature) $\partial^2 \nu/\partial B^2$. 
\textbf{(c)} Estimated Hahn-echo coherence time $T_2$ obtained from the rapid $T_2$ framework. 
The arrows indicate the Cartesian field directions $B_x$, $B_y$, and $B_z$. 
}
\label{fig:angle_sweep}
\end{figure*}

\renewcommand{\figurename}{FIG.}
\begin{figure}[ht]\centering\includegraphics[width=0.95\columnwidth]{ 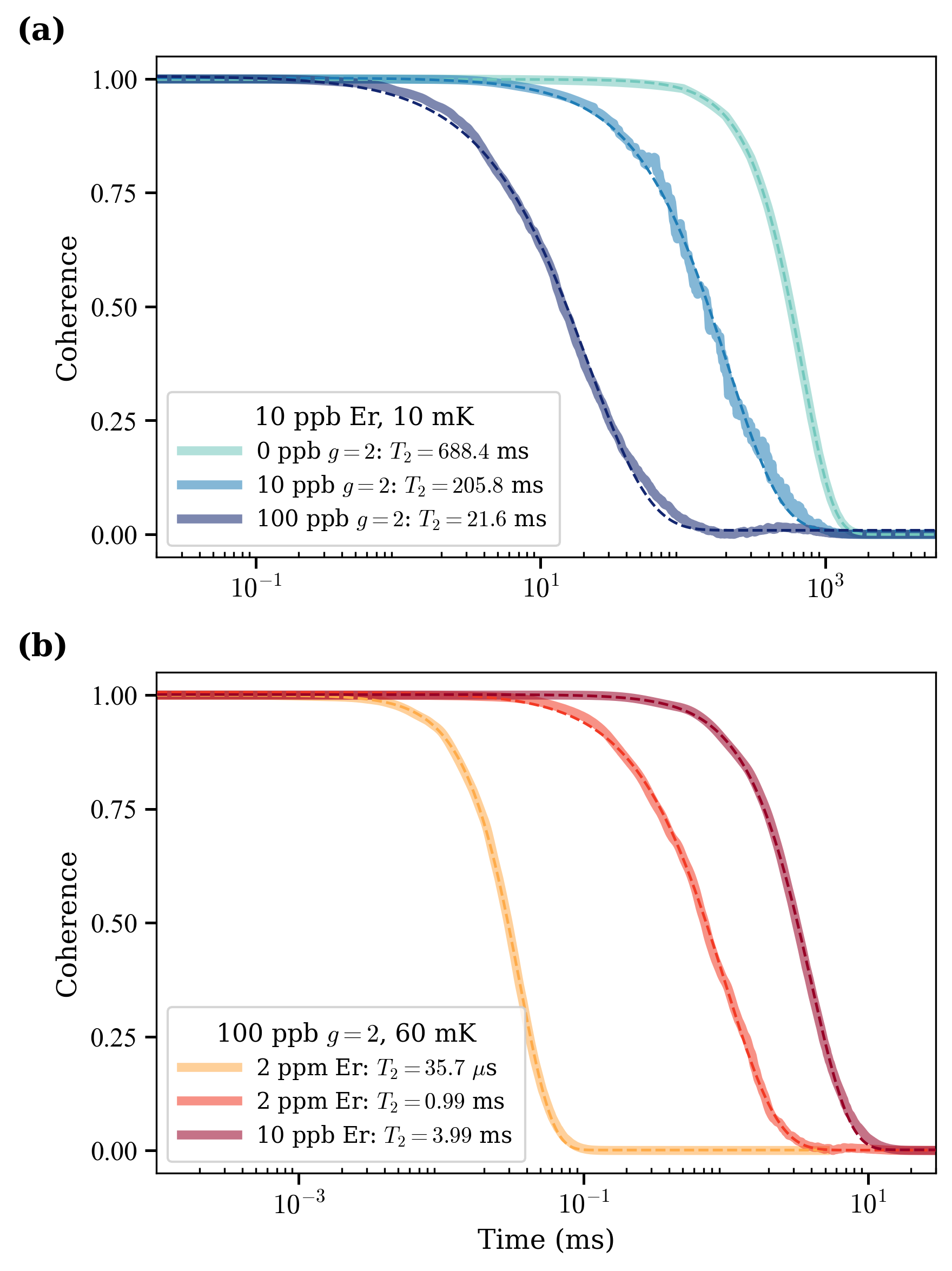}
\caption{{\textbf{Effect of additional low-$g$ paramagnetic defects on Hahn-echo coherence near clock transition 1.} (a) CCE-simulated Hahn-echo coherence decay $L(t)$ at $B_0=183$~G for a 10~ppb Er sample at 10~mK. Curves represent the nuclear-spin bath alone (0~ppb $g=2$; light blue), and in the presence of additional isotropic $g\approx2$ paramagnetic defect spins at 10~ppb (medium blue) and 100~ppb (dark blue) concentrations. (b) Coherence decay for a 100~ppb $g\approx2$ defect bath at 60~mK. The yellow curve corresponds to the transition studied by Seth \textit{et al.} for a 2~ppm Er sample. The light red and dark red curves correspond to conditions near clock transition 1 for 2~ppm and 10~ppb Er concentrations, respectively. Solid lines show the simulated coherence, and dashed lines show stretched-exponential fits used to extract the corresponding $T_2$ values.}}
\label{fig:g2_bath}
\end{figure}

\section{Hahn-Echo Coherence Near the Identified Clock Transitions}
\label{appendix:hahn_echo}

For completeness, we also evaluate the coherence near the identified clock transitions under the Hahn-echo sequence. At ultralow doping concentration and sub-kelvin temperature, decoherence is dominated primarily by the ${}^{17}\mathrm{O}$ nuclear spin bath ($I=5/2$), which generates slowly fluctuating local magnetic fields at the central $\mathrm{Er}^{3+}$ electron spin. As the Hahn-echo sequence partially refocuses quasi-static bath fluctuations, it provides a useful probe of the residual magnetic-field sensitivity near the clock points.

Figures.~\ref{fig:2}(a)-(c) show the calculated Hahn-echo coherence decay $L(t)$ as a function of the applied magnetic field $B_0$ near the three identified clock transitions, labeled transition 1, transition 2, and transition 3. As the magnetic field approaches the clock field, the coherence is strongly enhanced. Away from the clock condition, the coherence decays more rapidly due to the increased sensitivity of the transition frequency to nuclear spin fluctuations. Near the clock point, where the first-order frequency gradient is strongly suppressed, the decay becomes substantially slower and the coherence extends to longer times.

To quantify the statistical variation arising from different bath realizations, we extract the Hahn-echo coherence time $T_2$ by fitting each coherence trace $L(t)$ to a stretched exponential form for 150 independent nuclear spin bath configurations at each magnetic field value. The resulting distributions are shown in Figures.~\ref{fig:2}(d)-(f) as heat maps of $T_2$ versus $B_0$. The color scale indicates the number of bath realizations yielding a given coherence time, while the circles denote the ensemble-averaged $T_2$ values. For all three transitions, the ensemble-averaged $T_2$ remains strongly enhanced near-clock fields, confirming that suppression of first-order
magnetic susceptibility directly enhances coherence under realistic
many-body bath dynamics.

Consistent with the Ramsey results presented in the main text, these Hahn-echo simulations further confirm that the clock transitions identified from the central spin Hamiltonian correspond to optimal operating points for enhanced coherence in a realistic nuclear spin environment. The agreement between the Hahn-echo coherence maxima obtained from the full CCE simulations and the optimal points predicted by the rapid $T_2$ estimation therefore provides a direct validation of the Hamiltonian-based screening approach.

\section{Angular Dependence of Magnetic-Field Sensitivity}
\label{appendix:angular_dependence}

In addition to magnetic-field magnitude, the coherence properties near a clock transition can, in general, depend on the orientation of the applied field. In systems with anisotropic spin and field coupling, this leads to a strong angular dependence of the transition frequency $\nu(\mathbf{B})$ and its magnetic susceptibility. In the present case, however, $\mathrm{Er}^{3+}$ in $\mathrm{CeO}_2$ resides in a cubic crystal environment, for which the effective $g$-factor and hyperfine interaction are well approximated as isotropic. As a result, the transition frequency depends predominantly on the magnitude of the applied magnetic field, with only weak residual dependence on its orientation. To quantify this effect, we evaluated the angular dependence of the coherence properties at a magnetic-field magnitude chosen near one of the identified clock transitions. For each field direction on the unit sphere, we compute the first- and second-order derivatives of the transition frequency, as well as the corresponding coherence time $T_2$, using the rapid $T_2$ estimation framework described in Sec.~\ref{rapid_T2_estimation}. The resulting spherical maps are shown in Fig.~\ref{fig:angle_sweep}.

Figure~\ref{fig:angle_sweep}(a) and (b) show the first- and second-order magnetic-field sensitivities, $\partial \nu/\partial B$ and $\partial^2 \nu/\partial B^2$, respectively. Both quantities exhibit only weak variation over the sphere, indicating that the magnetic susceptibility is nearly isotropic. Consistent with this, the estimated coherence time $T_2$ shown in Fig.~\ref{fig:angle_sweep}(c) is also largely uniform across all field orientations.
Overall, these results demonstrate that, in contrast to many anisotropic spin systems, magnetic-field alignment does not play a significant role in determining coherence near clock transitions in $\mathrm{Er}^{3+}:\mathrm{CeO}_2$. This isotropic behavior is a direct consequence of the cubic crystal symmetry and further simplifies experimental implementation, as precise field orientation is not required to access optimal coherence.

\section{Effect of Additional \texorpdfstring{$g=2$}{g=2} Paramagnetic Spins on Coherence}
\label{appendix:effect of g=2 spins}
To assess the effect of extrinsic paramagnetic defects, we include dilute isotropic $g\approx2$ spin bath in the CCE simulations. Figure~\ref{fig:g2_bath}(a) shows the Hahn-echo coherence, approximately 2~G away from the first identified clock transition ($B_0=183$~G) with varying $g=2$ spin concentrations. As shown in the top panel, for a 10~ppb erbium concentration at 10~mK, the nuclear-spin bath alone yields $T_2=688.4$~ms. This coherence decreases to $205.8$~ms upon introducing a 10~ppb $g\approx2$ bath, and falls further to $21.6$~ms at 100~ppb. The Figure~\ref{fig:g2_bath}(b) examines the coherence with a 100~ppb $g\approx2$ bath at 60~mK and varying Er doping. For a non clock point ($\ket{m_I=-1/2}$ sub-level, at 495 Gauss) studied by Seth \textit{et al.} \cite{Seth2025} in a sample with 2~ppm erbium and 100~ppb  $g\approx2$ spins, we obtain a coherence time of $35.7$~$\mu$s, which is close to the experimentally reported value of $38.8$~$\mu$s. Under identical doping and temperature conditions near clock transition 1, we obtain an extended coherence of $0.99$~ms. If we further decrease the erbium concentration to 10~ppb near clock transition 1, the coherence time reaches $3.99$~ms. All curves are fitted with a stretched exponential function with a stretching factor ranging between 1 and 2, with the extracted $T_2$ values shown in Fig.~\ref{fig:g2_bath}.

\vspace{0.5in}

\bibliographystyle{apsrev4-2}
\bibliography{ref.bib}

\end{document}